\documentstyle[aps,twocolumn,epsf]{revtex}
\begin{document}

\draft

\title{Low temperature dynamics and laser-cooling of two-species
Coulomb chains for quantum logic}
\author{G. Morigi and H. Walther}
\address{Max Planck Institut f\"ur Quantenoptik, D-85748 Garching, Germany}
\date{\today}

\maketitle

\begin{abstract}
We study from the point of view of quantum logic
the properties of the collective oscillations of a linear chain of ions trapped in a linear Paul trap
and composed of two ion species. We discuss extensively
sympathetic cooling of the chain and the effect of anharmonicity on 
laser-cooling and quantum-information processing.
\end{abstract}



\section{Introduction}

The rapid development of trapping techniques for neutral and charged particles has constituted
a breakthrough in the investigation of quantum mechanical systems \cite{Traps}. Among the many interesting
experiments, ordered structures of charged ions have been achieved in Paul 
and Penning traps \cite{Walther2,Bollinger}. 
Such structures are composed from few up to thousands of particles,
and they originate at low temperature from the combined effect of the Coulomb
repulsive interaction among the ions and the trapping potential \cite{Dubin}. Therefore, 
their geometry depends intrinsically on the trap set-up.\\
\indent The field of interest of these so-called 
"Coulomb Crystals" is rather broad, 
and in quantum optics they find an application in the ion-trap quantum computer \cite{Zoller}. 
Here, a string of ions is proposed as a system for processing information, using two stable or 
metastable internal states of the ions for storing the quantum information and coherent 
interaction of the internal degrees of freedom of the single ions with the laser light
for generating the unitary operations which process the information, while the  
coupling among ions is provided by the collective vibrational excitations of the chain.
Present schemes for quantum information processing are based on the harmonicity
of the ionic motion \cite{JamesQG,Molme99}. This regime can be achieved by laser cooling \cite{Reviews}
the string of  
ions, and to a good extent the ions can be considered to 
vibrate harmonically around their equilibrium positions.\\
The coupling to the environment gives rise to phenomena which destroy the quantum coherence required
for information processing: Decoherence affects both the internal
and the motional quantum states. For motional states, decoherence
can be inhibited by applying laser-cooling to the ion-chain
on a regular rate or even continuously. Laser-cooling does not destroy the 
quantum information stored in the internal states, 
provided only some ions are addressed by the cooling laser ({\it cooling ions}), while the chain
is sympathetically cooled via the Coulomb interaction. Furthermore it allows for simultaneous 
information processing on the other ions ({\it qu-bits}), 
provided the quantum gates do not require quantum coherence among the 
vibrational levels, as for the gate proposed in \cite{Molme99} and realized in \cite{4ions}. 
In this respect, it is rather difficult to find a candidate ion which is, at the same time,
a good qu-bit and cooling ion. In addition, some realizations of ion strings for quantum information
do not rely on the spatial resolution of the ions with the laser \cite{4ions}.
Hence, one of the most
recent issues in the ion-trap quantum computer is to use two different ionic species which compose
the ion chain, one for quantum information processing, the other for laser-cooling \cite{MPQ,Zurek}.
This type of crystal, which we call here the {\it two-species Coulomb crystal}, has been 
already used for imaging the mechanical effect of radiation pressure on the crystallized ions and
for sympathetic cooling
of big ordered ionic structures \cite{Aarhus0}.\\
In this work we study the mechanical motion in a two-species linear crystal
from the point of view of quantum logic, illustrating
the general features and the differences from a linear crystal composed of 
ions of equal masses, and we discuss in some detail 
sympathetic cooling of the chain. The issue of decoherence is also discussed in 
connection to the mechanical properties of the system. In particular, in this paper we 
consider the decoherence due to the coupling of the ionic motion to the fluctuations of the
electric field (which we assume to have an instantaneous value with zero spatial 
gradient)
along the crystal, and thus couples with the center of mass motion. Finally, we
discuss the harmonic approximation and the effect of the anharmonic corrections on
laser cooling and quantum information processing. An analogous analysis has been 
presented in \cite{Zurek},
where they studied one specific scheme for sympathetic cooling.
Here, we study the more general system and its properties,
and seek schemes which are more suitable for realizing quantum logic.
In particular, we provide some examples calculated with $^{115}$In$^+$ and $^{25}$Mg$^+$ ions, which have
been currently trapped and cooled in Garching \cite{Peik99,Lange}.\\ 
The paper is organized as follows. In Section II the small oscillations formalism is introduced,
and the mechanical properties of a chain of ions are discussed in detail. In particular, 
sympathetic cooling of the two-species linear chain is studied, and the rates
of cooling for different crystal configurations are derived.
In Section III the effect of the anharmonic corrections on cooling and quantum logic
is discussed. Summary and conclusions are presented in Section IV.

\section{Small oscillations}

In this section we investigate the collective vibrations around the equilibrium position 
of a chain of ions confined in a linear Paul trap, 
following the lines of the literature of small oscillations
\cite{Goldstein}. We assume the motion to be 
one-dimensional, {\it i.e.} confined along the trap axis of the linear Paul trap.
This corresponds to assuming a very steep radial potential \cite{3D}. \\
\indent We consider a 
one--dimensional string of $N$ ions with charge $e$ and mass equal either to 
$M$ or $m$, which are aligned along the $\hat{x}$ axis, which corresponds to the axis of 
the linear Paul trap. Indicating with $i$ the 
position of the ion in the chain ($i=1,...,N$), the sequence of ionic
masses is described by the array ${\bf m}=(m_1,...,m_N)$ with $m_i=M,m$.
The ions are confined by the electrostatic potential $V_S$ and 
interact via the Coulomb repulsion \cite{Footnote1}. Sufficiently far away from the 
electrodes, $V_S$ can be considered harmonic and the total potential $V$ has the form:

\begin{equation}
\label{Vpot}
V=\sum_{i=1}^N \frac{1}{2}u_0x_i^2
+\frac{1}{2}\sum_{i=1}^N \sum_{j=1,j\neq i}^N\frac{e^2/4\pi\epsilon_0}{|x_i-x_j|}.
\end{equation}
where $x_i$ is the coordinate of the ion $i$ and $u_0$ is a constant with the
dimensions of an energy over a distance squared.
If the ions are sufficiently cold \cite{Dubin}, they crystallize around the 
classical equilibrium positions $x_i^{(0)}$, which are the solutions
of the set of equations $\partial V/\partial x_i|_{x_i^{(0)}}=0$. Those solutions
are independent of the mass, as the potential of the electrodes
interacts only with the ionic charges.
A characteristic quantity is the equilibrium distance between two ions:

\begin{equation}
\label{EqDistance}
x_0=x_2^{(0)}-x_1^{(0)}=\left(\frac{2e^2/4\pi\epsilon_0}{u_0}\right)^{1/3},
\end{equation}
where the ions are displaced symmetrically with respect to the center of the 
trap. This quantity scales the inter-ionic distance for a $N$ ion chain
\cite{JamesSteane}.\\
\indent Assuming that the collective motion of the ions around the equilibrium position is harmonic,
we approximate $V$ with its Taylor expansion around $x_i^{(0)}$ truncated to second order.
The dynamics of the system are described by the Lagrangian

\begin{equation}
\label{Laxial}
L=\frac{1}{2}\left[\sum_{i=1}^Nm_i\dot{q}_i^2-\sum_{i=1}^N V_{ij}q_iq_j\right]
\end{equation}
where $q_i=x_i-x_i^{(0)}$ are the displacements of the ions  from the equilibrium positions, 
and $V_{ij}$ are real coefficients which have the form:

\begin{eqnarray}
\label{Vij}
V_{ij}
&=&\frac{\partial^2}{\partial x_i\partial x_j}V(x_1,...,x_N)|_{\{x_i^{(0)}\}}\\
&=&u_0+2\sum_{k=1,k\neq i}\frac{e^2/4\pi\epsilon_0}{|x_i^{(0)}-x_k^{(0)}|^3}
~~~\text{~if~}i=j\nonumber\\
&=&-2\frac{e^2/4\pi\epsilon_0}{|x_i^{(0)}-x_j^{(0)}|^3}
~~~\text{~if~}i\neq j.\nonumber
\end{eqnarray}
From (\ref{Laxial}) the equations for the normal modes of the motion 
are

\begin{equation}
\sum_{j=1}^N V_{ij}\beta_j^{\alpha}=\lambda_{\alpha} m_i \beta_i^{\alpha}
\mbox{~~~with~~~}\alpha=1,...,N
\label{Sec0}
\end{equation}
where the eigenvalues $\lambda_{\alpha}$ are real given the hermiticity of $V_{ij}$, and
where ${\bf \beta^{\alpha}}$ is eigenvector at $\lambda_{\alpha}$.
Stable and harmonic
oscillations exist if the condition $\lambda_{\alpha}>0$ is fulfilled for any 
$\alpha$, as it occurs in this case. 
Under this condition the frequency of the normal mode $\Omega_{\alpha}$ is 
$\Omega_{\alpha}=\sqrt{\lambda_{\alpha}}$.
The eigenvectors $\beta_i^{\alpha}$
are orthogonal in the Riemannian metric with metric 
tensor ${\bf M}$, where ${\bf M}$ is a diagonal matrix whose diagonal corresponds 
to the array ${\bf m}$. For $\lambda_{\alpha}>0$,
introducing the mass-weighted coordinates $q_i'=\sqrt{m_i}q_i$ the eigenvalue 
problem can be rewritten as

\begin{equation}
\sum_jV_{ij}' {\beta_j^{\alpha}}'=\Omega_{\alpha}^2 {\beta_i^{\alpha}}'~~{\text{for}}~~
\alpha=1,...,N,
\label{Secular}
\end{equation}
where now $V_{ij}'=V_{ij}/\sqrt{m_im_j}$, and the metric tensor is the identity matrix, 
as for cartesian orthogonal coordinates. The eigenvalue problem is now equivalent to the one 
of $N$ identical ions of unitary mass. The matrix $\{{\beta_i^{\alpha}}^{\prime}\}$ 
defines
an orthogonal transformation, which reduces the system to the principal axes $\pi_{\alpha}$
of $V_{ij}$ and of the kinetic term: $\pi_{\alpha}=\sum_{i}{\beta_i^{\alpha}}^{\prime}q_i'$. In this 
representation the Lagrangian describes a set of $N$
independent harmonic oscillators with frequencies $\Omega_{\alpha}$.
We quantize the motion by associating a quantum mechanical oscillator with each mode. 
Then, denoting $a_{\alpha}$, $a_{\alpha}^{\dagger}$ 
the annihilation, creation operators for the mode $\alpha$, respectively, 
the coordinate $\pi_{\alpha}$ associated with the oscillator of frequency $\Omega_{\alpha}$ is written as
$\pi_{\alpha}=\sqrt{\hbar/2\Omega_{\alpha}}\left( a_{\alpha}+
a_{\alpha}^{\dagger} \right)$.
Going back to the original set of oblique coordinates $q_i$, they have the quantized form

\begin{equation}
q_i=\frac{1}{\sqrt{m_i}}\sum_{\alpha}\left({\beta}_i^{\alpha\prime}\right)^{-1}
\sqrt{\frac{\hbar}{2\Omega_{\alpha}}}
\left( a_{\alpha}+a_{\alpha}^{\dagger} \right).
\label{x_i}
\end{equation}

\indent Some general features can now be recognized.
From Eq. (\ref{Secular}) it is evident that the eigenmodes 
depend on the values of the ionic masses. Furthermore, since the matrix $V_{ij}$ is symmetrical 
by exchange of any pair of ions, the properties of the motion will be mainly characterized by the 
symmetries of the sequence ${\bf m}$. These properties are reflected in the eigenmodes of the 
motion $\{{\bf \beta_{\alpha}}\}$, and thus they affect the coupling of the crystal to radiation.\\
We discuss these points below. First, we consider 
the properties connected to two different values of the ionic masses, by analysing the case of two
 ions. Then, we discuss the ones connected to the symmetries of ${\bf m}$ by considering a 
three-ion crystal. Finally, on the basis of Eq. (\ref{x_i}) we study the mechanical effect of 
radiation on the crystal, and in particular sympathetic cooling of the chain.

\subsection{Two ions of different masses}

We analyse here the two-ion crystal, where the ions have masses $m$ and $M=\mu m$ 
with $\mu$ real parameter, $\mu>1$. For this case 
the secular equation (\ref{Secular}) can be solved analytically, and 
the eigenfrequencies of the motion have the form:

\begin{equation}
\label{Eigenfreqs}
\Omega_{\pm}^2=\frac{u_0}{m}\left(1+\frac{1}{\mu}\pm\sqrt{1
+\frac{1}{\mu^2}-\frac{1}{\mu}}\right)
\end{equation}
with corresponding displacements:

\begin{equation}
\label{Eigenvectors2}
q_{\pm}=N_{\pm}\left(\frac{1-\mu\mp\sqrt{1+\mu^2-\mu}}{\sqrt{\mu}},
\frac{1}{\sqrt{\mu}}\right)
\end{equation}
where the first and second components refer to the particles of mass $m$ and $M$, 
respectively.
Here, $N_{\pm}$ are the normalization factor, according to the scalar product
$q_i(1)q_j(1)+\mu q_i(2)q_j(2)=\delta_{ij}$ with $i,j=\pm$.\\
In Fig. 1(a) and (b) we plot the eigenfrequencies and the eigenvectors $q_{\pm}$, 
respectively, as a function of $\mu$. The ratio $\mu=1$ corresponds to the 
the well-known case of two ions of equal masses 
in a linear trap, where the ratio of the eigenfrequencies are
in the relation $1:\sqrt{3}$. For this value of $\mu$, $\Omega_-$ and 
$\Omega_+$ correspond to the 
center of mass (COM) and stretch mode frequencies, respectively, as it can also
be verified from Eq. (\ref{Eigenvectors2}). As $\mu$ increases, the value of the
eigenfrequencies decreases and tends asymptotically to the values $\Omega_-\to 0$ and 
$\Omega_+\to \sqrt{2u_0/m}$. The limit $\Omega_-\to 0$ corresponds to the
case where both ions stand still at their equilibrium position, as it can be seen in
Fig. 1(b), while for the limit $\Omega_+\to\sqrt{2u_0/m}$ 
the heavy ion does not move, and the light ion oscillates around its equilibrium
position.\\
In the following we will concentrate on the case $\mu>1$.
As it is apparent from Fig. 1 (b) and Eq. (\ref{Eigenvectors2}) the two modes preserve some
characteristics of the case of two ions with equal masses: 
In the mode of eigenfrequency $\Omega_-$ the ions
oscillate in phase, whereas for $\Omega_+$ they oscillate with opposite phases.
The two modes, however, do not correspond to the
COM and relative motion any longer. This can be understood by observing that,
in absence of interactions, the trap frequency for an ion of mass $m$ is proportional to
$1/\sqrt{m}$. This argument applies to an $N$-ion chain of two (or more) 
species, and can be verified by substituting the vector 
$q^{\rm COM}=(1,1,...,1)/\sqrt{N}$ describing the center of mass motion
inside the secular equation (\ref{Sec0}); one obtains 
$\sum_j V_{ij}=u_0=m_i \lambda^{\rm COM}$ for $i=1,...,N$, which cannot be
fulfilled for any value of $\lambda^{\rm COM}$, unless all masses $m_i$ are equal.\\
\indent The non-separability of the modes into center of mass and relative motion 
has some consequences on the dynamics. For example, for two ions the anharmonicity ({\it i.e.} 
the corrections to the harmonic approximation of the potential in Eq. (\ref{Laxial})) 
couples the two modes, 
whereas in the crystals of ions with equal masses the COM motion
is an exact eigenmode of the problem.
A further consequence is the coupling of both modes to the 
fluctuations of the electric field at the trap-electrodes, since none of the modes is
orthogonal to the COM motion. The strength with which each mode couples to this source
of decoherence is a function of the mass-ratio between the species $\mu$, 
as has been discussed in \cite{Zurek}.
 
\subsection{$N$-ion crystal}

As discussed above, 
the characteristic properties of the motion of a two-ion crystal are a function of the 
mass ratio $\mu$. For crystals with $N>2$ ions, some further parameters characterize
the properties of the motion:
The number of ions of each species and the sequence in which they are arranged.
These two features are described through the array ${\bf m}$. 
In one dimension, the relevant symmetry property of the sequence ${\bf m}$  
is the symmetry under reflections with respect to the
center of the trap (which is also the center of the string). This corresponds to an
invariance of the Hamiltonian under spatial parity transformations. 
Be $\Pi^{(N)}$ the parity operator, defined on the 
wave functions $|\phi(x_1,x_2,...,x_N)\rangle$ of the N-ion Hilbert space as 
$\Pi^{(N)}|\phi(x_1,x_2,...,x_N)\rangle=|\phi(-x_1,-x_2,...,-x_N)\rangle$. This operator
has eigenvalues $p=+1$ (even), $p=-1$ (odd), corresponding to the states with even and odd parity, 
respectively. If the array ${\bf m}$ is symmetric under reflections, the Hamiltonian
for the small oscillations commute with $\Pi^{(N)}$ and the eigenmodes of 
Eq. (\ref{Secular}) are also eigenvectors of $\Pi^{(N)}$ at  
the eigenvalue either $p=1$ or $p=-1$. From a simple evaluation of the number of degrees
of freedom, one can verify that 
the even modes are $N/2$ for an even number $N$ of ions and $(N-1)/2$ for $N$ odd.
In particular, for symmetry reasons the central ion does not move in the modes of
even parity of a chain with an odd number of ions.
Thus, these modes are independent of
the mass of the central ion, as it can be deduced from Eq. (\ref{Sec0}).\\
\indent It is instructive to take a closer look at the eigenfrequencies of a chain of $N$ ions
as a function of all possible sequences ${\bf m}$. We discuss the case of 3 ions, 
since it shares some similarities with the normal modes of a triatomic molecule, as discussed
in textbooks \cite{Goldstein}, and it exhibits features which can be extended to chains of large $N$.\\
In Fig. 2 we plot the eigenfrequencies of a $N=3$ chain as a function of all possible sequences
of Indium and Magnesium ions, where the sequences have been ordered with increasing total mass $M_C$ of the crystal. 
The mode of frequency $\Omega_1$ (solid line) is characterized by the oscillation in phase of the three ions. For 
ions of equal mass it corresponds to the center of mass mode, but in all cases still represents the
mechanical response of the whole crystal to excitations: In fact,
$\Omega_1\propto 1/\sqrt{M_C}$ and in general it does not show an appreciable
dependence on the order in which the ions are arranged. \\
The properties of the higher excitations depend on $\mu$ and on the sequence. In particular, the distance
among the eigenenergies changes depending on where the heavy ions are placed in the sequence.
In addition, symmetric sequences preserve the properties of a chain of three ions of equal masses. Thus,
the eigenmode of frequency $\Omega_2$ (dashed line) is characterized by the out-of-phase oscillation of the
external ions, whereas the central ion stands still. Hence,
$\Omega_2$ takes the same value for the sequences A and C, and for the sequences D and F. 
In asymmetric sequences the external ions still oscillate out-of-phase, but the oscillation 
amplitude of the central ion is large, independently of its mass, provided that $\mu\neq 1$.\\
In the symmetric sequences of the eigenmode $\Omega_3$ the external ions oscillate
in phase, whereas the central ion is out-of-phase, and its
amplitude is a monotonic function of $1/\mu$. For the sequence C, in particular,
$\Omega_2<\Omega_3$. The two frequencies are almost degenerate since both modes correspond to the case
where the outer ions move symmetrically with respect to the center, since the central ion in 
$\Omega_3$ has small displacements.
In asymmetric sequences (B,E) the light ions have large oscillation amplitudes,
whereas the displacements of the heavy ions are smaller.\\
\indent These properties have some immediate implications for quantum logic with a two-component chain.
For example, the eigenmodes of even parity of symmetric sequences are decoupled from
the fluctuations of the electric field at the electrodes, and thus  
are good candidate for the quantum bus. This issue have been discussed quantitatively for a particular 
sequence in \cite{Zurek}.\\
Furthermore, in symmetric sequences the parity operator $\Pi^{(N)}$ commutes with the 
one-dimensional Hamiltonian $H$: Hence, the eigenmodes with 
odd parity are not coupled via anharmonicity
to the eigenmodes with even parity. The axial motion, however, is coupled to the radial motion by
non-linearities, and in three-dimension there are no groups of modes which are decoupled from the others.
We discuss this point in section III. \\
Another important implication regards the spacing between the eigenfrequencies. In fact, 
when choosing a mode for quantum logic, the distance in energy among the eigenfrequencies 
should be taken into account, since the presence of quasi-degeneracies will lower the efficiency of single
mode addressing in quantum logic operations. Thus, it is not convenient to use
the mode with eigenfrequency $\Omega_3$ in the sequence C
as quantum bus, given its closeness to $\Omega_2$.\\
Finally, from the mechanical properties of the chain we can infer in which position 
the cooling ions should be placed for optimal sympathetic cooling of the chain. 
We analyse this aspect in the next subsection.

\subsection{Interaction of the crystal with light}

In the dipole approximation 
the coupling of the external degrees of freedom of an atom with radiation is
represented by the kick operator $\exp(i{\bf k}\cdot{\bf r})$, where ${\bf k}$ is the
wave vector of light and ${\bf r}$ the atomic position. 
In a Coulomb crystal of atomic ions, 
optical light couples to the internal degrees of freedom of a single
ion, which we consider here a two-level dipole transition, and to the external 
degrees of freedom of the collective motion. Thus, assuming that the ion $j$ scatters a 
photon and that $|\psi_i\rangle$ is the motional state of the crystal before the scattering, 
after the scattering the motion of the crystal is described by the state $|\psi_f\rangle$ given by:

\begin{eqnarray}
|\psi_f\rangle
&=& \exp(i k x_j)|\psi_i\rangle \nonumber\\
&=& \mbox{e}^{i\phi}\exp(ik q_j)|\psi_i\rangle
\label{kick}
\end{eqnarray}
where $\phi=k x_j^{(0)}$ is a real scalar, and $k$ is the projection of ${\bf k}$ on the
axis of the crystal. The coupling of
radiation to the collective modes of the crystal is visible by substituting 
Eq. (\ref{x_i}) into (\ref{kick}). Thus, the kick operator can be written as

\begin{equation}
\exp(i k q_j)=\Pi_{\alpha=1}^N\mbox{e}^{i\eta_j^{\alpha}
(a^+_{\alpha}+a_{\alpha})}
\end{equation}
where $\eta_j^{\alpha}$ is the Lamb-Dicke parameter
for the mode $\alpha$ and the ion $j$, and is defined as:  

\begin{equation}
\label{LambDicke}
\eta_{j}^{\alpha}=k\beta_j^{\alpha\prime}
\sqrt{\frac{\hbar}{2m_j\Omega_{\alpha}}}
\end{equation}
Through the Lamb-Dicke parameter  we can infer the mechanical response of the 
crystal to the scattering of a photon. For one ion of mass $m$, it corresponds to 
the square root of the ratio between the recoil frequency $\omega_R=\hbar k^2/2m$ and 
the trap frequency $\Omega$: $\eta=\sqrt{\omega_R/\Omega}$,
and it determines how many motional levels are
coupled by the scattering of one photon. The Lamb-Dicke regime corresponds to the case in 
which $\omega_R\ll\Omega$, and mathematically to the condition $\eta\sqrt{n}\ll 1$, where
$n$ is the vibrational state number. In this limit
the kick operator can be expanded in powers of $\eta$, and a change in the quantum 
motional state due to the incoherent scattering of one photon
is of higher order in $\eta$ \cite{Stenholm}. In this regime an ion can be sideband cooled
to the ground state of the vibrational motion \cite{Monroe}.\\
In presence of more than one ion, the Lamb-Dicke parameter $\eta_j^{\alpha}$ describes
how the displacement of the ion $j$ couples to the mode $\alpha$.
In particular, it determines {\it (i)} the possibility of addressing a single motional 
sideband, which appears when 
scanning a probe beam through the resonance frequency (and thus of exciting one
mode selectively) \cite{PRAIbk}, and {\it (ii)} the possibility
of laser-cooling a mode to its vibrational ground state, in analogy to the one-ion case.\\
Let us first consider the response to light of a two-ion crystal, and compare the case where 
the ions have equal masses, 
as discussed in \cite{PRAIbk}, with the case where they have different masses. 
In the first case,
due to the symmetry of the configuration the Lamb-Dicke parameters for each ion are equal (apart
for some difference in the sign). In the second case, we can see in Eq. (\ref{LambDicke})
that the Lamb-Dicke parameters depend on the mass of the ion, and given the asymmetry of the 
crystal the geometrical factor $\beta_j^{\alpha\prime}$ has different values 
for the two ions. One might be tempted to think that the Lamb-Dicke 
regime can be more easily accessed by addressing the heavier ions. 
This, however, is actually not true for all 
modes. This consideration is particularly applicable to the mode of lowest frequency, which shares
some properties with the center of mass motion, and in general describes the response of the crystal
as a whole to the mechanical excitation. For this mode, the displacements of the two ions
are comparable, and actually the displacement of the heavier ion is slightly larger, as 
can be seen in Fig. 1(b). On the contrary, the eigenmode with frequency $\Omega_+$ is characterized
by smaller displacements of the heavy ion than of the lighter one. For this mode, the heavy ion
might be well in the Lamb-Dicke regime, while the light ion is not.\\ 
This situation can be visualized by comparing the absorption spectra obtained by shining light
on each ion separately. We define the absorption
spectrum in a two-level transition with resonant frequency $\omega_0$ and driven by a laser
of frequency $\omega_L$ through 
the function $I(\delta)$, where $\delta=\omega_L-\omega_0$ is the detuning, 
which is evaluated by summing all contributions to
laser--excited transitions at frequency $\omega_L$ \cite{PRAIbk}:

\begin{equation}
\label{reson}
I(\delta)=\sum_{E_{\bf n}-E_{\bf l}=\delta}
|\langle {\bf n}|\exp(ikq_{j})|{\bf l}\rangle|^2 
P({\bf n})
\end{equation}
Here $|{\bf n}\rangle=|(n_1,n_2)\rangle$ are the motional states of energy $E_{\bf n}
=n_1\hbar\Omega_1+n_2\hbar\Omega_2$, with $n_{\alpha}$ 
occupation of the mode $\Omega_{\alpha}$ ($\alpha=1,2$), and 
$P({\bf n})$ is a normalized distribution over the motional states 
$|{\bf n}\rangle$. The coordinate $q_j$ characterizes the driven ion. In Fig. 3 
we plot $I(\delta)$ as a function of $\delta$ for a crystal composed of a Magnesium and an 
Indium ion.
Figure 3 shows the absorption spectrum when driving separately the Magnesium 
and the Indium ion, for a thermal distribution over the motional states
with average total energy $E=5\hbar\Omega_-$, where for the chosen parameters
$\Omega_-=0.552$ MHz and $\Omega_+=1.456$ MHz. In both cases the motional sidebands of the mode with
frequency $\Omega_-$ are visible, whereas when driving the In$^+$ ion
the sidebands of $\Omega_+$ almost disappear. 
In fact, as evaluated from (\ref{LambDicke})
the Lamb-Dicke parameters for $\Omega_-$ are $\eta_{\rm Mg}^{-}=0.22$,
$\eta_{\rm In}^{-}=0.38$, whereas the ones for $\Omega_+$ are $\eta_{\rm Mg}^{+}=0.5$, 
$\eta_{\rm In}^{+}=-0.06$, {\it i.e.} the weight of the motional sidebands for the mode 
$\Omega_+$ in Indium is two orders of magnitude smaller than the ones for the mode $\Omega_-$. 
It is interesting to compare this result with the case of two ions of equal masses. In that case 
the absorption spectrum is the same independently of which ion of the chain is driven. 
Then, if the COM mode is in the 
Lamb-Dicke regime, the relative motion mode is also in the Lamb-Dicke regime, since in that 
case the Lamb-Dicke
parameter scales simply as the inverse of the squared root of the eigenfrequency. \\
\indent For crystals with $N>2$ ions the factor 
$\beta_j^{\alpha}$ in (\ref{LambDicke})
reflects the structure of the chain and consequentely how the driven ion couples to 
the mode to cool. It thus contains some information on where the cooling ions should be placed
in the sequence so as to achieve more efficient cooling. This can be theoretically illustrated by
a rate equation, which describes cooling of one mode in a chain in the Lamb-Dicke
regime  \cite{footnote}. Here, we define the Lamb-Dicke regime with the condition:

\begin{equation}
\label{Bediengung}
\mbox{max}_{\{j\}}|\eta_{j}^{\alpha}|^2n_{\alpha}\ll 1 \mbox{~~~for~~~}\alpha=1,...,N
\end{equation}
where $n_{\alpha}$ denotes the occupation of the mode $\alpha$ and where the set of ions $\{j\}$ 
represents the set of positions of the cooling ions in the chain.
Assuming that the laser interacts with the internal two-level transition of the individual atom,
in second order perturbation theory in the parameter $g/\gamma$, with $g$ Rabi frequency and $\gamma$
decay rate of the atomic transition,
the excited state can be eliminated, and one obtains a set of equations projected on the
electronic ground state,
where populations and coherences between different motional states are coupled. 
In the Lamb-Dicke limit such coupling can be neglected,
thus reducing the equation for the $N$-ion density matrix $\rho$
in the low saturation limit to rate equations \cite{PRAIbk}. Furthermore, introducing 
the reduced density matrix $\rho_{\alpha}$ for the mode $\alpha$ 
defined by:

\begin{equation}
\rho_{\alpha}=\sum_{n_{\beta_1}}...\sum_{n_{\beta_{n-1}}}
\langle n_{\beta_1},...,n_{\beta_{n-1}}|\rho |n_{\beta_1},...,n_{\beta_{n-1}}
\rangle,
\end{equation}
with $\beta_1,...,\beta_{n-1}\neq\alpha$, one can derive the rate 
equation for cooling of the mode $\alpha$ in one-dimension \cite{Tesi}:  

\begin{eqnarray} 
\label{Eq4:cool_1mode}
\frac{\text{d}}{\text{d}t}P(n_{\alpha})
&=&\frac{g^{2}}{\gamma}{\sum_{\{j\}}}{\eta_j^{\alpha}}^2
[(n_{\alpha}+1)A_-^{\alpha}P(n_{\alpha}+1)\\
&-& ((n_{\alpha}+1)A_+^{\alpha}+n_{\alpha}A_-^{\alpha})
P(n_{\alpha})+n_{\alpha}A_+^{\alpha}P(n_{\alpha}-1)\nonumber
\end{eqnarray} 
the coupling with the other modes being of higher order in the Lamb-Dicke parameter.
Here, $P(n_{\alpha})=\langle n_{\alpha}|\rho_{\alpha} |n_{\alpha}\rangle$ and
the coefficients $A_{\pm}^{\alpha}$ are defined as 

\begin{eqnarray}
A_{+}^{\alpha}=\frac{1}{16\Omega_{\alpha}^2/\gamma^2+1}
+\frac{2}{5}\frac{1}{4\Omega_{\alpha}^2/\gamma^2+1},\\
A_{-}^{\alpha}=1
+\frac{2}{5}\frac{1}{4\Omega_{\alpha}^2/\gamma^2+1}.
\end{eqnarray}
where we have assumed that the Rabi coupling is spatially constant over the whole crystal
and equal to $g$, and that the laser is tuned on the first red sideband of the mode
$\Omega_{\alpha}$. Eq. (\ref{Eq4:cool_1mode}) is the sum of all contributions to cooling of mode 
$\alpha$ from the coupling $g$ of the cooling laser to the driven ions. It is fully equivalent 
to the equation for cooling of one ion in a harmonic trap of frequency $\Omega_{\alpha}$
in the Lamb--Dicke regime \cite{Stenholm}, apart from the scaling factor multiplying the 
term on the RHS of (\ref{Eq4:cool_1mode}):

\begin{equation}
\label{CoolRate}
W_{\alpha}=\sum_{\{j\}}|\eta_j^{\alpha}|^2
\end{equation}
Provided that (\ref{CoolRate}) is different from zero, it does not affect the steady state,
but scales the rate of cooling of the mode $\alpha$. The factor
$W_{\alpha}$ represents the contribution of the cooling ions in the array to the speed
of the process. Obviously, the largest cooling 
rate is achieved when all ions are driven by the cooling laser. In that case, 
$W_{\alpha}$ has the form $W_{\alpha}^{\rm max}=\sum_{j=1}^N|\eta_j^{\alpha}|^2
=\omega_R/\Omega_{\alpha}$, where $\omega_R$ is the recoil frequency of the single ion. 
The rate of cooling of each mode scales according to the relation 
$W_{\alpha}^{\rm max}=W_{\alpha'}^{\rm max}\Omega_{\alpha'}/\Omega_{\alpha}$, and it scales 
with the mass $m$ of the cooling ion as $W_{\alpha}^{\rm max}\propto \sqrt{1/m}$ 
since $\omega_R\propto 1/m$ and $\Omega_{\alpha}\propto 1/\sqrt{m}$.\\
For quantum logic we are interested in employing only some ions of the chain for
cooling. We look thus for the best sequence and mode for cooling, where for ``best sequence''
we intend a compromise between the highest number of ions for quantum logic ({\it i.e.} the
lowest number of cooling ions) and the best cooling rate.\\
In Fig. 4 we plot the factor $W_{\alpha}$ {\it vs} all possible
sequences of $N=3$ ions, made up of Magnesium and/or Indium ions, where the cooling ions are in (a)
Indium and (b) Magnesium ions. Note that between the curves in (a) and (b) there is a 
scaling factor corresponding, as expected, to the squared root of the ratio of the ionic masses. 
The curves in Fig. 4 can be easily interpreted by considering the 
properties of the modes, as discussed in the      
previous subsection. Thus, as the mode of frequency $\Omega_1$ is characterized by 
an oscillation in
phase of all ions, and does not strongly depend on the sequence,
the cooling rate increases
as the number of cooling ions increases. On the other hand, the mode with 
frequency $\Omega_2$ is 
mainly characterized by the oscillation of opposite phase of the ions placed 
externally. Thus, large
rates of cooling are achieved when the cooling ions are placed in the external positions.
In particular, the cooling rates of the sequences A and C,
and of the sequences D and F are equal. In fact, these configurations are symmetric under reflection and
the mode with eigenfrequency $\Omega_2$ has even parity. Thus, the central ion does not contribute
to cooling nor to quantum information processing with the mode. 
Finally, in symmetric configurations, the mode of 
frequency $\Omega_3$ is characterized by large oscillations of the central ions, and a relatively 
large rate of cooling is obtained by simply placing the cooling ion in the center, as
can be noticed for sequence C in Fig. 4(a) and sequence D in Fig. 4(b). In asymmetric
sequences (B,E) the rate of cooling in (a) is small, whereas in (b) is large,
as expected from the considerations made in the previous subsection.\\
For the case here considered, and on the basis of the mechanical properties only,
the best sequence of cooling is E provided that the cooling ion has lighter mass (Fig. 4(b)).\\
\indent In general, we can conclude that by preparing certain sequences one can have efficient sympathetic cooling
of some modes using a relatively small number of cooling ions. 
This characteristic does not depend strongly on the mass-ratio $\mu$ between the qubit and the cooling ions.
In a crystal with large total mass $M_C$, however, the Lamb-Dicke regime condition can be accessed more easily.
In this respect, it would be better to use heavier ions for quantum logic.\\
Finally, sequences with an even number $N$ of ions are to be preferred over sequences with odd $N$, so that
all positions in the chain contribute either to quantum logic or to the cooling process.

\section{Effects of the anharmonicity on cooling and quantum logic}

The harmonic approximation of the mechanical potential in (\ref{Vpot}) 
relies on the assumption that $V$ possesses strict local minima, around which
the motion is well-localized. In that case, the higher orders of its
Taylor expansion are a small correction. For two ions those terms have the form (for $x_2>x_1$):
\begin{equation}
\label{Anharmonic}
\delta V=\sum_{n=3}^{\infty}\delta V^{(n)}=
\sum_{n=3}^{\infty}(-1)^n\frac{e^2/4\pi\epsilon_0}{x_0^{n+1}}\left[q_2-q_1\right]^n
\end{equation}
where $\delta V^{(n)}$ is the $n$-th order correction. The effect of these terms, the so-called
anharmonicity, consists in causing shifts to the motional energies, and
mixing the eigenstates of the normal modes. Such mixing is in general a small correction to
the eigenstates of the ideal case, but it may become particularly enhanced because of 
quasi
degeneracies among the motional energies of the states. In fact, the density of 
motional states of an $N$-ion chain in the interval of energy $[E,E+\delta E]$ is 
approximately $D(E)\propto E^{N-1}$. Thus, as the number $N$ of ions
increases and/or for larger values of the motional energies, the dynamics of 
quasi-degenerate states are definetely affected by the anharmonicity.
Here, we discuss the effects of the departures from the ideal 
harmonic system, first on sideband cooling
and then on the efficiency of quantum logic gates. \\
\indent In sideband cooling 
the laser addresses the motional sideband of the mode to be cooled. 
Thus, cooling will be efficient as long as the shift in energy caused by the anharmonicity is much smaller
than the frequency of one phonon of the mode addressed. 
On the other hand, the mixing between the eigenstates will constitute a 
thermalization effect among the modes, and it will not constitute 
an obstacle to cooling as long as all modes are at sufficiently 
low temperatures, so that $\delta V$ is a small correction to the whole
system.\\
To obtain some estimates, we evaluate the order of magnitude of the 
shift to the energy in first order perturbation
theory, and ask for which range of values of the vibrational numbers sideband cooling may still work.
In \cite{PRAIbk} it has been shown that in the perturbative
regime $\delta V\approx \delta V^{(3)}$ \cite{footAnh}. Thus:

\begin{equation}
\label{deltaif}
\langle\delta V^{(3)}\rangle\approx \frac{e^2/4\pi\epsilon_0}{x_0}\left(\frac{a_{0\alpha}}{x_0}\right)^3
n_{\alpha}^{3/2}
\end{equation}
where $a_{0\alpha}=\sqrt{\hbar/m_i\Omega_{\alpha}}$, $n_{\alpha}$ vibrational number of the 
mode $\alpha$, and $m_i$ mass of the lighter ion. In deriving (\ref{deltaif}) 
we have assumed $n_{\alpha}/\Omega_{\alpha}\ge n_{\beta}/\Omega_{\beta}$ ($\alpha,\beta=1,2,...,N$). 
Taking two ions, one Indium and one Magnesium, $\Omega_{1}=1$MHz and the mass of Magnesium $^{25}$Mg$^{+}$, then 
$|\langle\delta V\rangle|/\hbar\approx 6\times 10^{-3}n_1^{3/2}\Omega_1$, which implies that first order 
perturbation theory holds for $n_{1}\ll 80$ ($n_2\ll 30$). In this limit, a laser
tuned on the first sideband to the red of the mode $\Omega_1$ cools the system to the ground state.
We have verified this conjecture numerically: We have taken a two-ion chain composed
of an Indium and a Magnesium ion, and considered sideband cooling in the Lamb-Dicke regime of 
one of the modes, comparing the case in which the mechanical potential is fully harmonic with 
the case where the third and fourth orders in the anharmonic expansion have been included.
We have not noticed any significant difference between the two cases, and the system is 
cooled efficiently to the ground state. In particular, no visible effect
could be interpreted as due to the anharmonic coupling between quasi-degenerate states. 
However, according to the above estimates, in the interval of states of the numerical calculation
the spectrum of energy is not very ``dense''. Then,
in order to verify numerically the effect of anharmonocities in presence of quasi-degeneracies
we take a system with exact degeneracies, and more specifically with
two modes of frequencies $\Omega_1=\Omega$ and $\Omega_2=2\Omega$. Here,
in the Lamb-Dicke regime a laser sideband-cools the mode of frequency 
$\Omega_1$. We compare the harmonic with the anharmonic case,
where here we simply substitute the chosen values $\Omega_1,\Omega_2$
in the quantized form of the displacements of Eq. (\ref{x_i}).
In Fig. 5 the average occupation number for the modes (a) $\Omega_1$ and (b) $\Omega_2$ is 
plotted as a function of the time. The dashed and solid lines correspond to the harmonic 
and anharmonic case, respectively.
In the harmonic case the mode $\Omega_1$ is cooled and $\Omega_2$ is like ``frozen'', 
since it is coupled to radiation at higher orders in the Lamb-Dicke parameter \cite{PRAIbk}.
In the anharmonic case the rate of cooling of the mode addressed is slowed down, whereas the mode 
$\Omega_2$ is simultaneously cooled: The system is cooled as a whole, but on a relatively slower
time-scale. Thus, the two modes thermalize on a time-scale which is faster than 
the cooling one.\\
\indent With respect to quantum information processing, 
present quantum logic schemes with ions are based on the harmonic properties 
of the motion. Thus, they are affected
both by the shift in energy and the mixing induced by the anharmonicity. Quantum gates which require the preparation
of the system in the ground state of the motion will be very weakly affected, since in that part of the spectrum 
the anharmonic perturbation is extremely small, and there are no
quasi-degeneracies. On the other hand, the perturbation may affect the efficiency of hot gates, since they
operate on higher-lying motional states. Thus, the speed of
hot gates must be faster than the rate of anharmonic coupling between quasi-degenerate states. 
We can define a time-scale
$\tau_{\rm Anh}$ for the anharmonic perturbation $\tau_{\rm Anh}\approx \hbar/|\langle\delta V\rangle|$. 
The typical duration of a quantum gate must be shorter than $\tau_{\rm Anh}$. In presence of 
degenerate states which are coupled by three-phonon transitions ({\it i.e.} for 
$\Omega_{\alpha}\sim 2\Omega_{\beta}$), from (\ref{deltaif})
$\tau_{\rm Anh}\sim 10~\mu$sec given $\Omega_{\alpha}=1$ MHz and $n_{\alpha}\sim 10$. This estimate,
which is rather worrying if compared with the duration of a quantum gate \cite{Molme99,Speed}, 
reflects the worst case, which might occur for certain sequences, large number
of ions and large excitations. Note that, if the degenerate states are coupled by a four-phonon 
transition ({\it i.e.} for $\Omega_{\alpha}\sim 3\Omega_{\beta}$),
then $\tau_{\rm Anh}\sim 600~\mu$sec. In general, we expect this problem to arise when 
working with a large number of ions and for high excitations. It could be minimized by choosing symmetric sequences:
in that case the ``effective density'' of motional states which are coupled by the anharmonicities of
the axial potential will decrease, since only states with the same parity will be coupled to each other,
whereas the coupling with the radial degrees of freedom is of higher order.\\
As a general rule, however, one should avoid degeneracies among the radial and the 
axial frequencies.

\section{Conclusions}

We have studied the small oscillations behaviour of a two-component linear crystal,
with particular emphasis on the applications to sympathetic cooling and quantum logic with the ions,
and have discussed the effect of anharmonicity on the operations of the ion-trap quantum computer.
We have seen that higher efficiency in quantum logic and sympathetic
cooling are achieved by selecting the right ionic sequences. That raises the issue of how to prepare the desired
sequence of ions. A rigorous investigation in this direction
should take into account the full non-linear potential in the
three-dimensional space and it is subject of on-going investigations.\\

\indent
We would like to thank P. Lambropoulos for many stimulating discussions
and the critical reading of this manuscript, and W. Lange, S. K\"ohler, V. Ludsteck, 
E. Peik, who are involved in the experimental realization of the ion structure
discussed here. This work is supported in parts by the European Commission within the
TMR-networks ERB-FMRX-CT96-0087 and ERB-FMRX-CT96-0077.


\begin{figure}
\begin{center}
\epsfxsize=0.5\textwidth
\epsffile{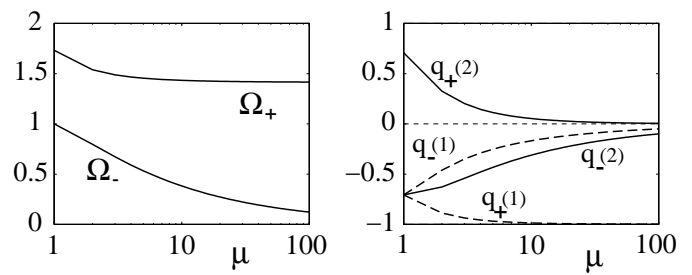}
\end{center}
\caption{(a) Eigenfrequencies of the two-ion crystal as a function of the mass
ratio $\mu$. The frequencies are rescaled by the factor $\sqrt{u_0/m}$. (b) 
Corresponding displacements in the oblique axes plotted as a function of $\mu$.
Solid line: particle of mass $M$. Dashed line:
particle of mass $m$. 
}
\end{figure}

\begin{figure}
\begin{center}
\epsfxsize=0.25 \textheight
\epsffile{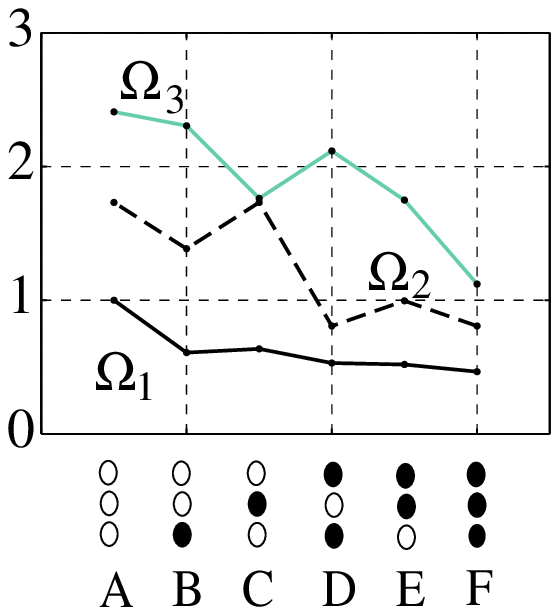}
\end{center}
\caption{Eigenfrequencies $\Omega_{\alpha}$ {\it vs}
all possible configurations of a $N=3$ ionic sequence made up of Indium (black circles)
and/or Magnesium (white circles) ions. The eigenfrequencies are rescaled by the value of $\Omega_1$ for
a chain of Magnesium ions.
}
\end{figure}

\begin{figure}
\begin{center}
\epsfxsize=0.4 \textwidth
\epsffile{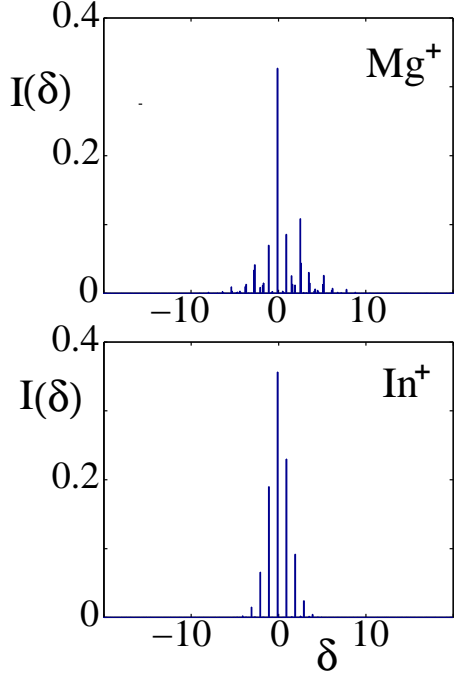}
\end{center}
\caption{Absorption spectrum $I(\delta)$ {\it vs} the detuning $\delta$ 
for a crystal of an Indium and Magnesium ion
obtained shining light on Magnesium and Indium, for a thermal distribution with
average energy $5\hbar\Omega_-$. Here, $\eta_{\rm Mg}^{-}=0.22$, 
$\eta_{\rm Mg}^{+}=0.5$,
$\eta_{\rm In}^{-}=0.38$, $\eta_{\rm In}^{+}=-0.06$, as follows from (\ref{LambDicke}).
}
\end{figure}

\begin{figure}
\begin{center}
\epsfxsize=0.25\textheight
\epsffile{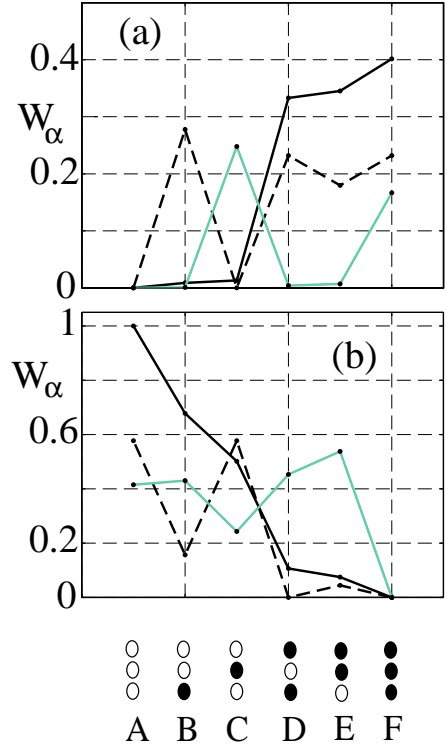}
\end{center}
\caption{Cooling rate $W_{\alpha}$ {\it vs}
all the possible configurations of a $N=3$ ionic sequence made up of Indium 
(black circles)
and/or Magnesium (white circles) ions, where $W_{\alpha}$ has been rescaled
by the value of $W_{\alpha}^{\text{max}}$ for a chain of Magnesium ions. Solid line:
Eigenmode $\Omega_1$; Dashed line: Eigenmode $\Omega_2$; Grey line: Eigenmode 
$\Omega_3$ [cif.
Fig. 3].
}
\end{figure}

\begin{figure}
\begin{center}
\epsfxsize=0.5\textwidth
\epsffile{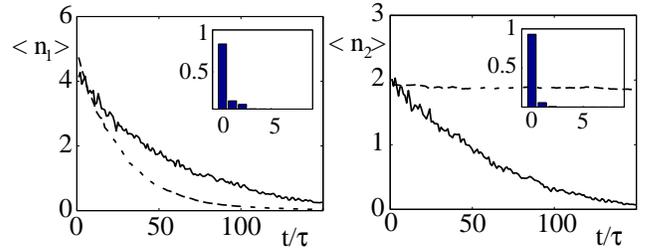}
\end{center}
\caption{
Average occupation of the mode (a) $\Omega_1=\Omega$ and (2) $\Omega_2=2\Omega$ as a function of the time in unit of
the optical pumping time $\tau$, for the harmonic (dashed line) and 
anharmonic case (solid line). Here, $\Omega=1$ MHz, $\eta_{\Omega}=0.1$, $\eta_{2\Omega}=0.1/\sqrt{2}$, 
$\gamma=0.1\Omega$, $\delta=\omega_L-\omega_0=-\Omega$, $g=0.01\Omega$, $\tau=\gamma/g^2$. 
Insets: Population of the modes (a) $\Omega_1$ and (b)
$\Omega_2$ as a function of their vibrational number at time $t=150\tau$ for the anharmonic case.
}
\end{figure}


\begin{thebibliography}{99}
\bibitem{3D}
The treatment here presented can be analogously extended to 
three-dimensions.
For a discussion on the transverse motion we refer the reader to \cite{Zurek}.
\bibitem{Footnote1}
Here, we neglect the dipole-dipole interaction and focus on the 
regime where the inter-ionic distance is considerably
larger than the wavelength of the light coupling to the internal transition.
\bibitem{footnote}
Note that ground state cooling can be achieved also in systems not strictly in the
Lamb-Dicke regime, when the selective laser excitation of only one sideband of the
motion is in general not possible (see \cite{PRAIbk}). On the other hand, 
present schemes for quantum
logic gates which are scalable to $N$ ions are stricly based on the Lamb-Dicke regime 
assumption, see \cite{JamesQG,Molme99}.
\bibitem{footAnh}
To be precise, the third order correction does not cause any energy shift in first order perturbation
theory. However, here we are interested just to the order of magnitude of the perturbation, and this
estimate suits to our purposes.
\end{thebibliography}
\end{document}